\documentclass[11pt]{article}
\usepackage{hyperref}
\usepackage{graphicx}
\usepackage{amsmath}
\usepackage{float}
\usepackage[a4paper, total={6in, 9in}]{geometry}
\usepackage[round]{natbib}  
\title{PyLongslit: a simple manual Python pipeline for processing of astronomical long-slit spectra recorded with CCD detectors}
\author{
    Kostas Valeckas\thanks{Niels Bohr Institute, Copenhagen University. ORCID: \href{https://orcid.org/0009-0007-7275-0619}{0009-0007-7275-0619}} \and
    Johan Peter Uldall Fynbo\thanks{Cosmic Dawn Center, Niels Bohr Institute, Copenhagen University. ORCID: \href{https://orcid.org/0000-0002-8149-8298}{0000-0002-8149-8298}} \and
    Jens-Kristian Krogager\thanks{Centre de Recherche Astrophysique de Lyon. ORCID: \href{https://orcid.org/0000-0002-4912-9388}{0000-0002-4912-9388}} \and
    Kasper Elm Heintz\thanks{Cosmic Dawn Center, Niels Bohr Institute, Copenhagen University. ORCID: \href{https://orcid.org/0000-0002-9389-7413}{0000-0002-9389-7413}}
}
\date{27 March 2025}

\begin{document}

\maketitle

\section*{Summary}

We present a new Python pipeline for processing data from astronomical long-slit spectroscopy observations recorded with CCD detectors. The pipeline is designed to aim for \textbf{simplicity}, \textbf{manual execution}, \textbf{transparency} and \textbf{robustness}. The goal for the pipeline is to provide a manual and simple counterpart to the well-established semi-automated and automated pipelines. The intended use-cases are \textbf{teaching} and \textbf{cases where automated pipelines fail}. For further elaboration, please see the \hyperref[sec:statement-of-need]{Statement of need}. \\

\noindent From raw data, the pipeline can produce the following output:
\begin{itemize}
    \item A calibrated 2D spectrum in counts and wavelength for every detector pixel.
    \item A 1D spectrum extracted from the 2D spectrum in counts per wavelength (for point-like objects).
    \item A flux-calibrated 1D spectrum in $\frac{\text{erg}}{\text{s} \cdot \text{cm}^2 \cdot \text{\AA}}$ (for point-like objects).
\end{itemize}

\noindent The products are obtained by performing standard procedures for detector calibrations \citep{Howell_2006, handbook}, cosmic-ray subtraction \citep{cr_1, cr_2}, and 1D spectrum extraction \citep{photutils, Horne_1986}.

\section*{Statement of need}
\label{sec:statement-of-need}

A natural approach when developing data-processing pipelines is to seek precision and automation. The trade-off for this is code complexity and "black-box" solutions, where the process of the pipeline is often masked, and the quality-assessment output is made under the assumption that the user knows how to interpret it. \\

\noindent In research, this is a reasonable trade-off, as a certain level of user skill and experience can be assumed. However, in a teaching paradigm, simplicity and transparency are often more favorable, even when this means loss of precision and automation. The PyLongslit pipeline is designed to rely on simple code and manual execution, supported by a large array of quality-assessment plots and extensive documentation. The algorithms are designed to produce research-quality results, yet while prioritizing simplicity over high precision. The reason for this is to create a robust and transparent pipeline, where every step of the execution is visualized and explained. We see this as being especially valuable in teaching scenarios and for users that are new to spectroscopic data processing. Furthermore, we hope that the simple coding style will invite users of all skill levels to contribute to the code. \\

\noindent An early beta version of the software was user-tested during the Nordic Optical Telescope\footnote{\url{https://www.not.iac.es/}} IDA summer-course 2024\footnote{\url{https://phys.au.dk/ida/events/not-summer-school-2024}}, where all student groups were able to follow the documentation and successfully process data without any significant assistance.\\

\noindent During the development of the software, it became apparent that the manual nature of the pipeline is also useful for observations where automated pipelines might fail. The PyLongslit pipeline can revert to manual methods instead of using mathematical modeling when estimating the observed object trace on the detector. This is especially useful for objects that have low signal-to-noise ratio or where several objects are very close to each other on the detector. Furthermore, extraction can be performed with either optimal extraction methods \citep{Horne_1986} or by summing detector counts for a box-like object shape \citep{photutils} (this can be useful for emission-line dominated objects).

\section*{Pipeline}

The figures below describe the pipeline structure. The inspiration for the pipeline architecture is taken from the very popular (but no longer maintained) IRAF \citep{IRAF}. In a broad sense, there are three stages of the data processing, all explained in separate figures. The diamond shapes in the figures represent different pipeline routines that are called directly from the command line, solid arrows are hard dependencies (must-have), dashed arrows are soft dependencies (can use), and the rectangles represent input files and pipeline products.

\begin{figure}[H]
    \centering
    \includegraphics[width=\textwidth]{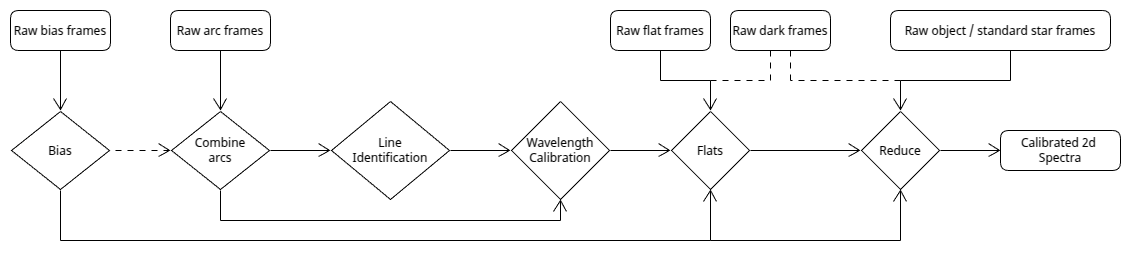}
    \caption{Step 1 - processing raw data. In this step, all the raw observation and calibration frames are used to construct calibrated 2D spectra. After this step, all procedures are performed directly on the calibrated 2D spectra, and the raw frames are no longer used.}
    \label{fig:raw_processing}
\end{figure}

\begin{figure}[H]
    \centering
    \includegraphics[width=\textwidth]{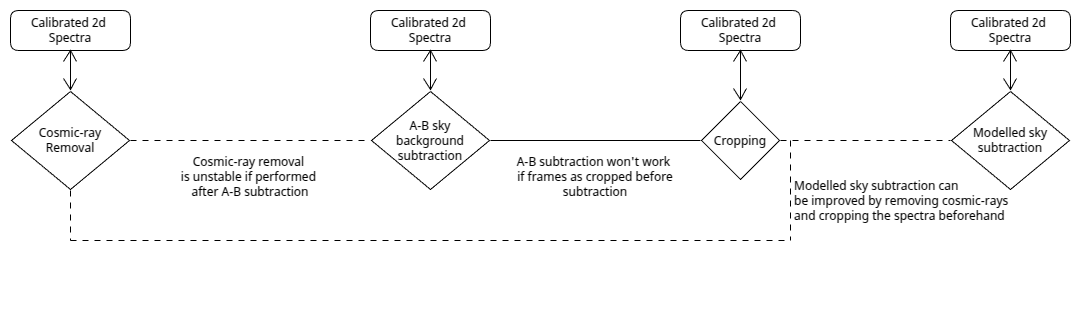}
    \caption{Step 2 - further processing of the calibrated 2D spectra. In this step, the user can deploy cosmic-ray removal, sky-background subtraction, and crop the spectra. All procedures alter the 2D spectra in place. All of the steps are optional, but there are some dependencies — these are described in the figure.}
    \label{fig:further_processing}
\end{figure}

\begin{figure}[H]
    \centering
    \includegraphics[width=\textwidth]{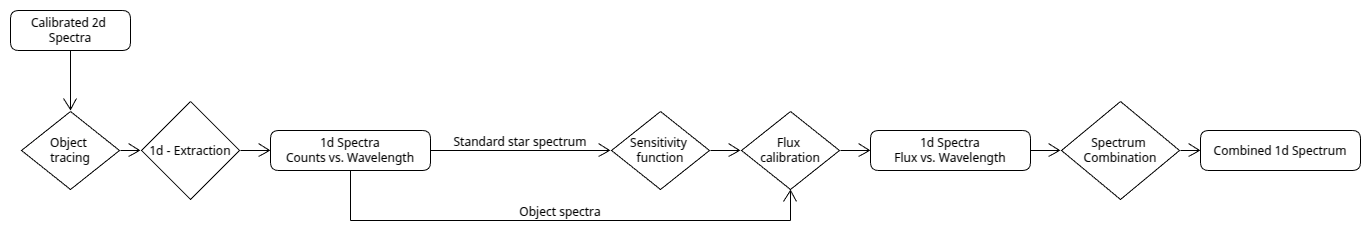}
    \caption{Step 3 - 1D spectrum extraction. In this step, objects are traced, extracted, flux calibrated, and combined (if several spectra of the same object exist).}
    \label{fig:1d_extraction}
\end{figure}

\noindent The pipeline is controlled by a configuration file that has to be passed as an argument to every pipeline procedure. The different parameters of the configuration file are described in the documentation\footnote{\url{https://kostasvaleckas.github.io/PyLongslit/index.html}}.

\section*{Evaluation}

To test the pipeline for correctness, we run the pipeline on data from two long-slit instruments: NOT ALFOSC\footnote{\url{https://www.not.iac.es/instruments/alfosc/}} and GTC OSIRIS\footnote{\url{https://www.gtc.iac.es/instruments/osiris/}}, and compare the results with the results from the well-established, semi-automated PypeIt Python pipeline \citep{pypeit:zenodo, pypeit:joss_arXiv, pypeit:joss_pub}.

\begin{figure}[H]
    \centering
    \includegraphics[width=\textwidth]{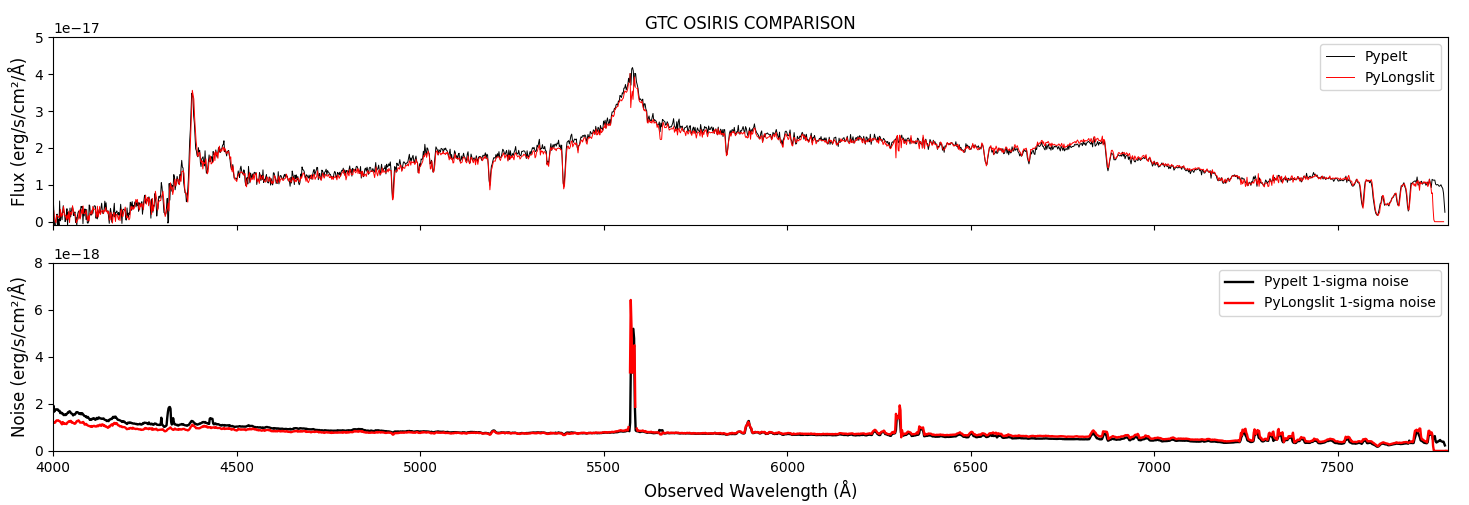}
    \caption{GTC OSIRIS observation of GQ1218+0823.}
    \label{fig:gtc}
\end{figure}

\begin{figure}[H]
    \centering
    \includegraphics[width=\textwidth]{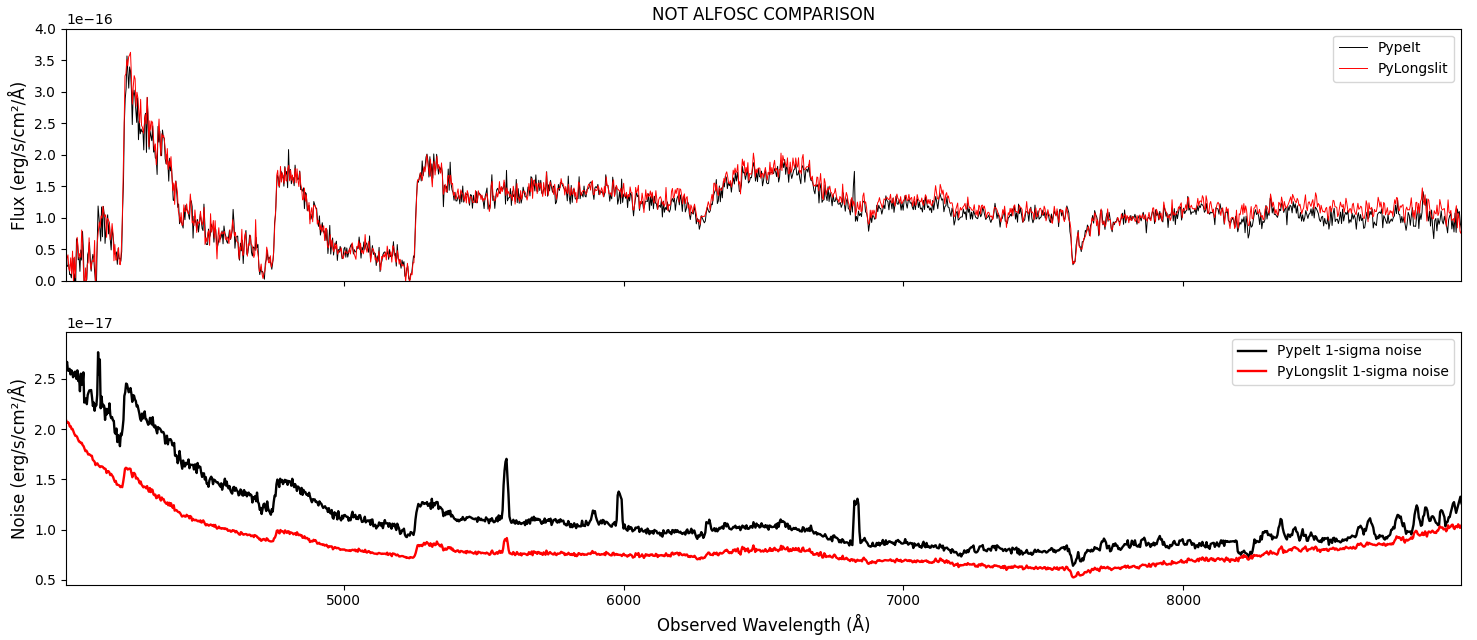}
    \caption{NOT ALFOSC observation of SDSS\_J213510+2728.}
    \label{fig:alfosc}
\end{figure}

\noindent We see good agreement between the two pipelines on both axes, for both the extracted spectrum and the noise estimation. For NOT ALFOSC data, we see some deviation in the error magnitude and error related to strong sky-lines. This is due to skipping modeled sky subtraction in the PyLongslit run, as the A-B sky background subtraction was sufficient by itself. We calculate the noise numerically for a cut of the spectrum where the flux is somewhat constant to confirm that the PyLongslit noise indeed is smaller in magnitude:

\begin{figure}[H]
    \centering
    \includegraphics[width=\textwidth]{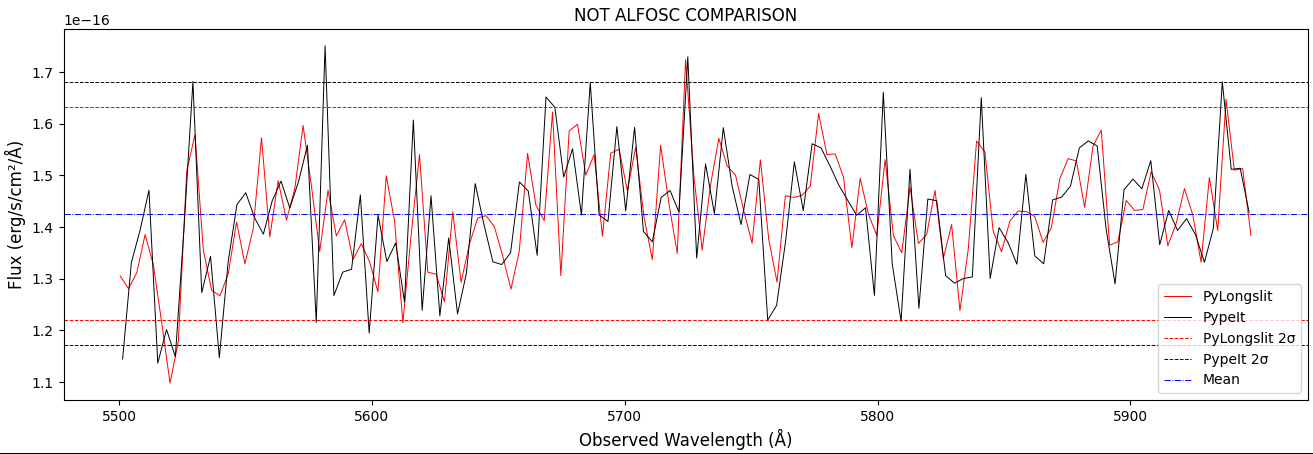}
    \caption{Numerical noise comparison for NOT ALFOSC observation of SDSS\_J213510+2728.}
    \label{fig:alfosc_zoom}
\end{figure}

\section*{Limitations}

As mentioned in the \hyperref[sec:statement-of-need]{Statement of need}, PyLongslit favors simplicity over high precision. Furthermore, the pipeline is designed to be \textbf{instrument independent}. Due to these design choices, the pipeline does not account for any instrument-specific phenomena, such as detector fringing and alike. The pipeline will likely be less precise than an instrument-specific pipeline (depending on the implementation of the latter). The code is written with a focus on \textbf{loose coupling}, and therefore the pipeline code can be used as a starting point for an instrument-specific pipeline.

\section*{Acknowledgements}

We thank the participants of the Nordic Optical Telescope IDA summer-course 2024 for very useful feedback on the software.

\bibliographystyle{plainnat}
\bibliography{paper}

\end{document}